\documentclass{PoS}

\title{Shaft Inflation and the Planck satellite observations}

\ShortTitle{Shaft Inflation}

\author{\speaker{Konstantinos Dimopoulos}
\\
Consortium for Fundamental Physics, Physics Department, Lancaster University,\\ 
Lancaster LA1 4YB, UK\\
        E-mail: \email{k.dimopoulos1@lancaster.ac.uk}}

\abstract{%
A new family of inflation models is introduced and studied. The models are
characterised by a scalar potential which, far from the origin, approaches an
inflationary plateau in a power-law manner, while near the origin becomes 
monomial, as in chaotic inflation. The models are obtained in the context of 
global supersymmetry starting with a superpotential, which interpolates from a 
generalised monomial to an O'Raifearteagh form for small to large values of the
inflaton field respectively. It is demonstrated that the observables obtained, 
such as the scalar spectral index, its running and the tensor to scalar ratio, 
are in excellent agreement with the latest observations, without any 
fine-tuning. Moreover, by widening mildly the shaft in field space, it is shown
that sizable tensors can be generated, which may well be observable in the 
near future.}

\FullConference{18th International Conference From the Planck Scale to the Electroweak Scale \\
		25-29 May 2015\\
		Ioannina, Greece }

\begin{document}

\section{Introduction}

The latest observations from the Planck satellite 
\cite{planck1,planck2,planck3} confirm the
vanilla predictions of cosmic inflation for the primordial curvature
perturbation in that the latter is predominantly Gaussian (non-Gaussianities 
have not been observed, with upper bound $f_{\rm NL}^{\rm local}=0.9\pm 5.7$), 
adiabatic (no isocurvature contribution has been observed, with upper bound to 
less that 3\%), statistically isotropic (no statistical anisotropy has been 
observed, with upper bound to less than 2\%) and almost scale-invariant, but 
with a significant red tilt ($n_s=0.968\pm0.006$). Moreover, the Planck data 
favour canonically normalised, single-field, slow-roll inflation \cite{planck1}.
In fact, in conjunction with other data, Planck seems to favour an inflationary
plateau \cite{bestinf}. 

There have been many examples of such inflationary
modes, such as the original $R^2$-inflation \cite{staro}, Higgs inflation 
\cite{higgs} or T-model inflation \cite{Tmodel}. However, most of these attempts
consider an exponential approach to the inflationary plateau. Here we design a 
model, which approaches the inflationary plateau in a power-law manner, 
offering distinct observational signatures.

\section{Bottom-up versus top-down approach}

In inflationary model-building one can identify two broad strategies.
The top-down scenario corresponds to designing models based on ``realistic'' 
constructions, for example inspired by string theory, supergravity etc. Then,
one looks for specific signatures in the data (e.g. non-Gaussianity). Since the
latest Planck data favour single-field, slow-roll inflation, they seem to 
support such relatively straightforward constructions. 

In contrast, the bottom-up scenario amounts to inflationary model constructions,
which are ``suggested'' by the data, i.e. they are data-inspire 
``guess-stimates''. As such, this approach uses the Early Universe as 
laboratory to investigate fundamental physics, in the best tradition of 
particle cosmology. We adopt this strategy (see also Ref.~\cite{shaft}). 
Our model proposes a power-law approach to the inflationary plateau in the 
context of global supersymmetry.

\section{The scalar potential for Shaft Inflation}

\begin{figure}
\begin{center}
\includegraphics[width=.6\textwidth]{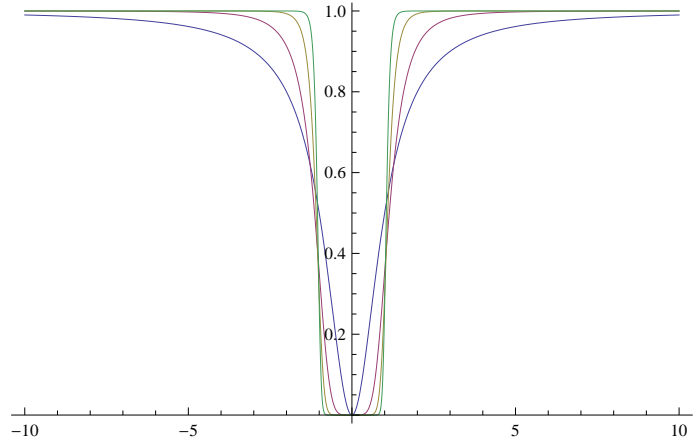}
\caption{%
The scalar potential in shaft inflation for \mbox{$n=2,4,8$} and $16$. The 
shaft becomes sharper as $n$ grows. Far from the origin the potential 
approximates the inflationary plateau with \mbox{$V\approx M^4$}. Near the 
origin the potential becomes monomial, as in chaotic inflation. 
}
\label{fig0}
\end{center}
\end{figure}

Consider a toy-model superpotential of the form: 
\mbox{$W=M^2\Phi^{nq+1}/(\Phi^n+m^n)^q$}, where $n,q$ are real numbers and
$M,m$ are mass-scales. For \mbox{$|\Phi|\gg m$}, this superpotential approaches
an O' Raifearteagh form \mbox{$W\simeq M^2\Phi$} leading to de~Sitter inflation.
For \mbox{$|\Phi|\ll m$}, the superpotential becomes \mbox{$W\propto\Phi^{nq+1}$}
leading to chaotic inflation. To simplify it even further, we choose to
eliminate the numerator, and take \mbox{$q=-1/n$}. We end up with the 
superpotential for Shaft Inflation \cite{shaft}:
\begin{equation}
W=M^2\left(\Phi^n+m^n\right)^{1/n}.
\label{W}
\end{equation}
To obtain the scalar potential, we consider \mbox{$\Phi=\phi e^{i\theta}$},
where $\phi,\theta$ are real scalar fields with $\phi>0$.%
\footnote{A normalisation factor of 
$1/\sqrt 2$ has been absorbed in the mass scales.}
Then the scalar potential is:
\begin{equation}
V=M^4|\Phi|^{2(n-1)}|\Phi^n+m^n|^{2(\frac{1}{n}-1)}=
\frac{M^4\phi^{2(n-1)}}{[\phi^{2n}+m^{2n}+2\cos(n\theta)m^n\phi^n]^{\frac{n-1}{n}}}.
\end{equation}
The potential is minimised when \mbox{$n\theta=2\ell\pi$}, with $\ell$ being an
integer. Further, noting that \mbox{$-\phi=\phi e^{i\pi}$}, we can make the
potential symmetric over the origin [\mbox{$V(\phi)=V(-\phi)$}] if 
\mbox{$n=2\ell$}, i.e. even. In this case,
\begin{equation}
V(\phi)=M^4\phi^{2n-2}(\phi^n+m^n)^{\frac{2}{n}-2},
\label{V}
\end{equation}
for all real values of $\phi$. %(icluding negative)
From the above we see that the scalar potential has the desired behaviour,
for \mbox{$n\geq 2$}, i.e. it approaches a constant \mbox{$V\approx M^4$}
for \mbox{$\phi\gg m$}, while for \mbox{$\phi\ll m$} the potential becomes 
monomial, with \mbox{$V\propto\phi^{2(n-1)}$}, see Fig.~\ref{fig0}. 

\vspace{-1cm}

\section{The spectral index and the tensor to scalar ratio}

\subsection{\boldmath Slow-roll parameters, $n_s$ and $r$}

From Eq.~(\ref{V}), we readily obtain the slow-roll parameters as
\begin{eqnarray}
& & \epsilon\equiv\frac12 m_P^2\left(\frac{V'}{V}\right)^2=
2(n-1)^2\left(\frac{m_P}{\phi}\right)^2\left(\frac{m^n}{\phi^n+m^n}\right)^2
\label{eps}\\
& & \eta\equiv m_P^2\frac{V''}{V}= 
2(n-1)\left(\frac{m_P}{\phi}\right)^2\left(\frac{m^n}{\phi^n+m^n}\right)
\frac{(2n-3)m^n-(n+1)\phi^n}{\phi^n+m^m},
\label{eta}
\end{eqnarray}
where the prime denotes derivative with respect to the inflaton field and
$m_P=2.4\times 10^{18}\,$GeV is the reduced Planck mass. Hence, 
the spectral index of the curvature perturbation is
\begin{equation}
n_s=1+2\eta-6\epsilon=1-4(n-1)\left(\frac{m_P}{\phi}\right)^2
\frac{m^n[(n+1)\phi^n+nm^n]}{(\phi^n+m^n)^2}.
\label{ns}
\end{equation}
To rewrite the above as functions of the remaining e-folds of inflation $N$ 
we have to investigate the end of inflation.
It is straightforward to see that inflation is terminated when 
\mbox{$|\eta|\simeq 1$} so that, for the end of inflation, we find
\begin{equation}
\phi_{\rm end}\simeq m_P\left[2(n^2-1)\alpha^n\right]^{1/(n+2)},
\label{fend}
\end{equation}
where we assumed that \mbox{$\phi>m$} (so that the potential deviates
from a chaotic monomial) and we defined
\begin{equation}
\alpha\equiv\frac{m}{m_P}\,.
\label{alpha}
\end{equation}
Using this, we obtain $\phi(N)$ 
\begin{eqnarray}
N=\frac{1}{m_P^2}\int_{\phi_{\rm end}}^\phi\frac{V}{V'}{\rm d}\phi & \simeq &
\frac{1}{2(n-1)(n+2)\alpha^n}\left[\left(\frac{\phi}{m_P}\right)^{n+2}-
\left(\frac{\phi_{\rm end}}{m_P}\right)^{n+2}\right]
\label{N}\\
& \Rightarrow & \phi(N)\simeq m_P\left[2(n-1)(n+2)\alpha^n
\left(N+\frac{n+1}{n+2}\right)\right]^{1/(n+2)}.
\label{fN}
\end{eqnarray}
Inserting the above into Eqs.~(\ref{eps}) and (\ref{ns})
respectively we obtain the tensor to scalar ratio $r$ and the spectral index 
$n_s$ as functions of $N$:
\begin{eqnarray}
& & r=16\epsilon=32(n-1)^2\alpha^{\frac{2n}{n+2}}\left[2(n-1)(n+2)
\left(N+\frac{n+1}{n+2}\right)\right]^{-2(\frac{n+1}{n+2})}
\label{r}\\
& & n_s=1-2\,\frac{n+1}{n+2}\left(N+\frac{n+1}{n+2}\right)^{-1}.
\label{nsN}
\end{eqnarray}
Notice that only $r$ is dependent on $m$ (through $\alpha$), which means that
$r$ can be affected by changing $m$ without disturbing $n_s$. We will return to
this possibility later.

\subsection{Examples}

To investigate the performance of the model, we consider the two extreme cases
for the values of $n$, namely \mbox{$n=2$} and \mbox{$n\gg 1$}. For 
illustrative purposes we take \mbox{$\alpha=1$}, i.e. \mbox{$m=m_P$}.

\subsubsection{\boldmath $n=2$}

In this case the scalar potential becomes
\begin{equation}
V(\phi)=M^4\frac{\phi^2}{\phi^2+m^2}.
\label{Vquad}
\end{equation}
We see that the above can be thought of as a modification of quadratic 
chaotic inflation, because after the end of inflation, the inflaton field 
oscillates in a quadratic potential. However, for large values of the inflaton 
the potential approaches a constant. %as implied by the Planck observations. 
This potential has been obtained also in S-dual superstring inflation 
\cite{Sdualinf} with \mbox{$\alpha=1/4$} and also in radion assisted gauge 
inflation \cite{RAGI} with \mbox{$\alpha\sim 10^{-3/2}$}. In this case,
Eqs.~(\ref{r}) and (\ref{nsN}) become
\begin{equation}
r=\frac{32\alpha}{\left[8\left(N+\frac34\right)\right]^{3/2}}
\qquad{\rm and}\qquad
n_s=1-\frac32\left(N+\frac34\right)^{-1}.
\end{equation}
From the above, we find the values for $n_s$ and $r$, as shown in 
Table~\ref{tab1}.
\begin{table}
\begin{center}
\begin{tabular}{|c|c|c|}\hline
$N$ & $n_s$ & $r$\\\hline\hline
50 & 0.970 & 0.0039\\
60 & 0.975 & 0.0030\\\hline
\end{tabular}
\end{center}
\caption{Values of $n_s(N)$ and $r$ in the case $n=2$.}
\label{tab1}
\end{table}

\subsubsection{\boldmath $n\gg 1$}

In the opposite extreme \mbox{$n\gg 1$}, Eqs.~(\ref{r}) and (\ref{nsN}) become
\begin{equation}
r=\frac{8\alpha^2}{n^2(N+1)^2}\rightarrow 0
\quad{\rm and}\quad
n_s=1-\frac{2}{N+1}\,.
\label{nbig}
\end{equation}
The spectral index is now the same as in the original $R^2$~inflation model
\cite{staro} (also in Higgs inflation \cite{higgs}), which is not surprising 
since we expect power-law behaviour to approach the exponential when 
\mbox{$n\rightarrow\infty$}. The values of $n_s$, in this case, are shown in 
Table~\ref{tab2}.

\begin{figure}
\begin{center}
\vspace{-12cm}

\mbox{\hspace{-3.4cm}
\includegraphics[width=1.4\textwidth]{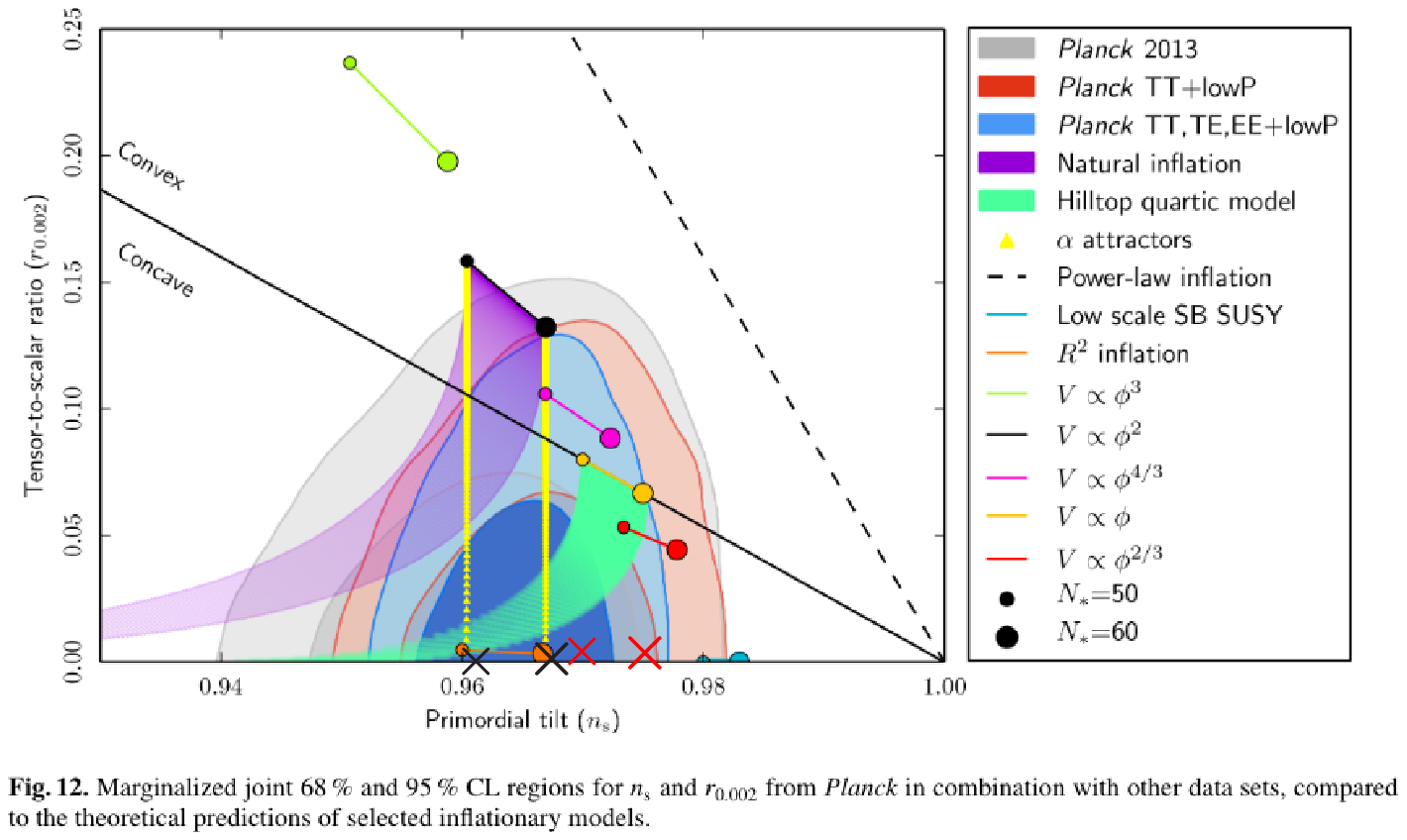}}
\vspace{-9.5cm}

\caption{%
Shaft inflation for \mbox{$n=2$} is depicted with the large \{small\} red cross
for \mbox{$N\simeq 60$} \{\mbox{$N\simeq 50$}\}. Shaft inflation for 
\mbox{$n\gg 1$} is depicted with the large \{small\} black cross for 
\mbox{$N\simeq 60$} \{\mbox{$N\simeq 50$}\}.
%The difference with $R^2$-inflation (orange dots) stems from taking into 
%account the contribution of $\phi_{\rm end}$. 
%so that \mbox{$N\rightarrow N+1$} in Eq.~(\ref{nbig}). 
Intermediate values of $n$ lie in-between the depicted points.
As evident, there is excellent agreement with the Planck observations.%
}
\label{fig1}
%\vspace{-1cm}
\end{center}
\end{figure}
\begin{table}
\begin{center}
\begin{tabular}{|c|c|}\hline
$N$ & $n_s$ \\\hline\hline
50 & 0.961 \\
60 & 0.967 \\\hline
\end{tabular}
\end{center}
\caption{Values of $n_s(N)$ in the case $n\gg 1$ (where $r\approx 0$).}
\label{tab2}
\end{table}
From the above, we find that the values for $n_s$ and $r$ are very close to the
best fit point for the Planck data for all values of $n$, 
as shown in Fig.~\ref{fig1}.\footnote{%
The crosses are superimposed to the original Planck paper image, taken from
Ref.~\cite{planck1}, which includes also the original caption.}

%\mbox{\hspace{1cm}}

%\pagebreak

\section{Gravitational waves}

\begin{figure}
\begin{center}
\vspace{-4cm}
\includegraphics[width=.9\textwidth]{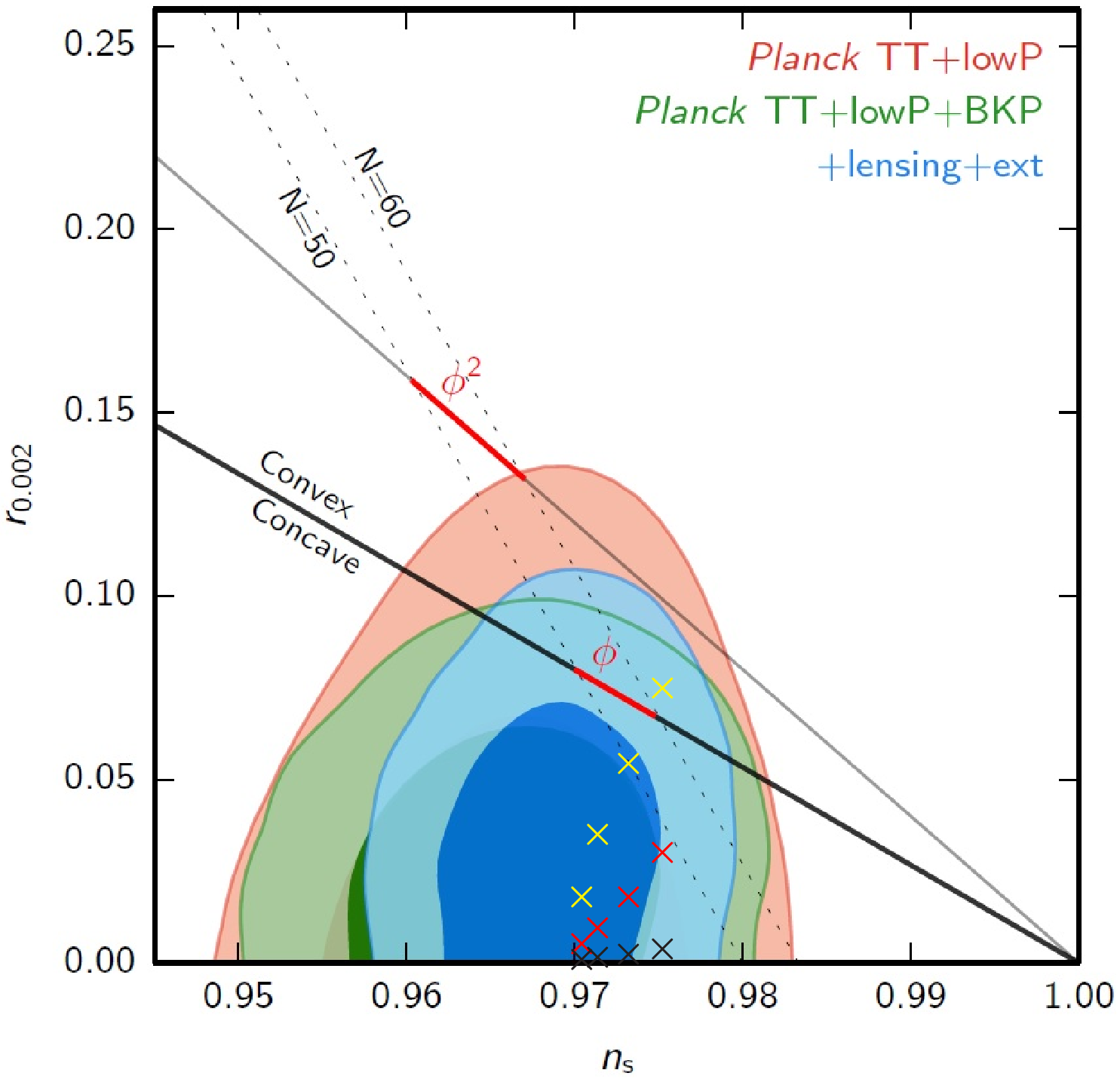}
\vspace{-6cm}
\caption{%
Shaft inflation predictions for \mbox{$N=60$}. The crosses in the image 
correspond to \mbox{$n=2,4,6,8$} as depicted from right to left. 
Black crosses correspond to \mbox{$\alpha=1$} (\mbox{$m=m_P$}),
red crosses correspond to \mbox{$\alpha=2\sqrt{8\pi}\approx 10$} 
(\mbox{$m=2M_P$}) and yellow crosses correspond to 
\mbox{$\alpha=5\sqrt{8\pi}\approx 25$} (\mbox{$m=5M_P$}).
It is evident that, for mildly super-Planckian values of $m$ the model 
predictions lie at the verge of observability.}
\label{fig2}
\vspace{-1cm}
\end{center}
\end{figure}

Planck observations, in conjunction with BICEP2 and Keck Array data suggest
\mbox{$r\leq 0.1$} \cite{PKB}. As we have seen, in Shaft Inflation, 
\mbox{$r\propto\alpha^{2n/(n+2)}$}, while there is no $\alpha$-dependence of 
$n_s$. Thus, by changing $m$, $r$ can vary without affecting the spectral index 
(c.f. Eq.~(\ref{alpha})). Therefore, sizeable tensors can be attained by 
widening the shaft in field space. Indeed, rendering $m$ mildly super-Planckian
can produce potentially observable values of $r$ as shown in Table~\ref{table3},
where Eq.~(\ref{alpha}) suggests that 
\mbox{$m=\alpha\,m_P=\frac{\alpha}{\sqrt{8\pi}}M_P$}, with 
$M_P=1.2\times 10^{19}\,$GeV being the Planck mass. 
\begin{table}
\begin{center}
\begin{tabular}{|c||c|c|c|c|}\hline
$n$ & $n_s$ & $r\;(\alpha=1)$ 
& $r\;(\alpha=2\sqrt{8\pi}\approx 10)$ 
& $r\;(\alpha=5\sqrt{8\pi}\approx 25)$ \\\hline\hline
2 & 0.975 & 0.0030 & 0.0299 & 0.0747\\
4 & 0.973 & 0.0008 & 0.0168 & 0.0570\\
6 & 0.971 & 0.0003 & 0.0089 & 0.0352\\
8 & 0.970 & 0.0001 & 0.0052 & 0.0227\\\hline
\end{tabular}
\end{center}
\caption{Values of $n_s$ and $r$ for $N=60$ and $n=2,4,6,8$. Three choices of 
$\alpha=m/m_P$ are depicted, which correspond to $m=m_P$, $m=2M_P$ and $m=5M_P$,
where \mbox{$M_P=\sqrt{8\pi}\,m_P$}.
It is shown that, with $m$ mildly super-Planckian, $r$ can approach the 
observational bound $r<0.1$ without affecting $n_s$.}
\label{table3}
\end{table}
Thus, we see that with 
$m\simeq 5\,M_P$ we can have $r\simeq 0.07$, which is on the verge of 
observability. This is shown clearly in Fig,~\ref{fig2}.

\section{More on Shaft Inflation}

The running of the spectral index is easily obtained as
\begin{equation}
\frac{{\rm d}n_s}{{\rm d}\ln k}=
-\frac{2\left(\frac{n+1}{n+2}\right)}{\left(N+\frac{n+1}{n+2}\right)^2}.
%\sim-\frac{2}{N^2}\sim - 10^{-3}
\end{equation}
In the two extreme cases, this gives
\begin{eqnarray}
n=2: & & 
\frac{{\rm d}n_s}{{\rm d}\ln k}=-\frac{3}{2\left(N+\frac34\right)^2}=
-4.064\times 10^{-4}\\
n\gg 1: & &
\frac{{\rm d}n_s}{{\rm d}\ln k}=-\frac{2}{\left(N+1\right)^2}=
-5.375\times 10^{-4} 
\end{eqnarray}
where the numerical values correspond to $N=60$. Thus, for all values of $n$,
we the above suggests:
\mbox{$\frac{{\rm d}n_s}{{\rm d}\ln k}\approx-(4-5)\times 10^{-4}$},
which is in agreement with the Planck findings:
\mbox{$\frac{{\rm d}n_s}{{\rm d}\ln k}=-0.003\pm0.007$}.

Finally, the inflationary scale is determined by the COBE constraint
\begin{equation}
\sqrt{{\cal P}_\zeta}=\frac{1}{2\sqrt 3\pi}\frac{V^{3/2}}{m_P^3|V'|},
\end{equation}
where \mbox{${\cal P}_\zeta=(2.208\pm 0.075)\times 10^{-9}$} is the spectrum of
the curvature perturbation \cite{planck2}. This provides an estimate for the 
required value of $M$
\begin{equation}
\left(\frac{M}{m_P}\right)^2=4\sqrt 3(n-1)\alpha^{-\frac{n}{n+2}}
\pi\sqrt{{\cal P}_\zeta}\left[2(n\!-\!1)(n\!+\!2)\!\left(N+\frac{n+1}{n+2}\right)
\right]^{-\frac{n+1}{n+2}}.
\label{M}
\end{equation}
For illustrative purposes, using 
\mbox{$\sqrt{{\cal P}_\zeta}\simeq 4.7\times 10^{-5}$}, \mbox{$n=2$}, 
\mbox{$\alpha=1$} (i.e. \mbox{$m=m_P$}) and \mbox{$N=60$} we find
\mbox{$M=7.7\times 10^{15}\,$GeV}, which is very near the scale of grand 
unification, as expected.

\section{Conclusions}

Planck data favour single-field, slow-roll inflation, characterised by a 
scalar potential which approaches an inflationary plateau. In contrast to many 
other successful models, Shaft Inflation approaches this plateau in a power-law
manner. Shaft Inflation is based on a simple superpotential: 
\mbox{$W=M^2\left(\Phi^n+m^n\right)^{1/n}$}. Without any fine tuning 
(\mbox{$m\sim m_P$} and \mbox{$M\sim 10^{16}\,$}GeV, i.e. the scale of grand 
unification) Shaft Inflation produces a scalar spectral index very close to the
Planck sweet spot with very small (negative) running, in agreement with Planck.
Rendering $m$ mildly super-Planckian one can easily obtain potentially 
observable tensors without affecting the spectral index. The challenge in now 
to obtain realistic setups which can realise the (deceptively) simple Shaft 
Inflation superpotential.

\section*{Acknowledgements}
KD is supported (in part) by the Lancaster-Manchester-Sheffield Consortium for 
Fundamental Physics under STFC grant ST/L000520/1.


\begin{thebibliography}{99}
\bibitem{planck1}
P.~A.~R.~Ade {\it et al.} [Planck Collaboration],
  %``Planck 2015 results. XX. Constraints on inflation,''
  arXiv:1502.02114 [astro-ph.CO].
  %%CITATION = ARXIV:1502.02114;%%

\bibitem{planck2}
P.~A.~R.~Ade {\it et al.} [Planck Collaboration],
  {\em Planck 2015 results. XIII. Cosmological parameters},
  arXiv:1502.01589 [astro-ph.CO];
  %%CITATION = ARXIV:1502.01589;%%

\bibitem{planck3}
P.~A.~R.~Ade {\it et al.} [Planck Collaboration],
  {\em Planck 2015 results. XVII. Constraints on primordial non-Gaussianity},
  arXiv:1502.01592 [astro-ph.CO].
  %%CITATION = ARXIV:1502.01592;%%

\bibitem{bestinf}
J.~Martin, C.~Ringeval, R.~Trotta and V.~Vennin,
  {\em The Best Inflationary Models After Planck},
  JCAP {\bf 1403} (2014) 039
  [arXiv:1312.3529 [astro-ph.CO]].
  %%CITATION = ARXIV:1312.3529;%%

\bibitem{staro}
A.~A.~Starobinsky,
  {\em A New Type of Isotropic Cosmological Models Without Singularity},
  Phys.\ Lett.\ B {\bf 91} (1980) 99;
  %%CITATION = PHLTA,B91,99;%%
%A.~A.~Starobinsky,
  %``The Perturbation Spectrum Evolving from a Nonsingular Initially De-Sitte r%Cosmology and the Microwave Background Anisotropy,''
  Sov.\ Astron.\ Lett.\  {\bf 9} (1983) 302.
  %%CITATION = SALED,9,302;%%

\bibitem{higgs}
F.~L.~Bezrukov and M.~Shaposhnikov,
  {\em The Standard Model Higgs boson as the inflaton},
  Phys.\ Lett.\ B {\bf 659} (2008) 703
  [arXiv:0710.3755 [hep-th]].
  %%CITATION = ARXIV:0710.3755;%%

\bibitem{Tmodel}
A.~Linde,
  {\em Inflationary Cosmology after Planck 2013},
  arXiv:1402.0526 [hep-th].
  %%CITATION = ARXIV:1402.0526;%%

\bibitem{shaft}
 K.~Dimopoulos,
  %``Shaft Inflation,''
  Phys.\ Lett.\ B {\bf 735} (2014) 75
  [arXiv:1403.4071 [hep-ph]].
  %%CITATION = ARXIV:1403.4071;%%

\bibitem{Sdualinf}
A.~de la Macorra and S.~Lola,
  {\em Inflation in S dual superstring models},
  Phys.\ Lett.\ B {\bf 373} (1996) 299
  [hep-ph/9511470].
  %%CITATION = HEP-PH/9511470;%%

\bibitem{RAGI}
M.~Fairbairn, L.~Lopez Honorez and M.~H.~G.~Tytgat,
  {\em Radion assisted gauge inflation},
  Phys.\ Rev.\ D {\bf 67} (2003) 101302
  [hep-ph/0302160].
  %%CITATION = HEP-PH/0302160;%%

\bibitem{PKB}
P.~A.~R.~Ade {\it et al.} [BICEP2 and Planck Collaborations],
  {\em Joint Analysis of BICEP2/$Keck  Array$ and $Planck$ Data},
  Phys.\ Rev.\ Lett.\  {\bf 114} (2015) 101301
  [arXiv:1502.00612 [astro-ph.CO]].
  %%CITATION = ARXIV:1502.00612;%%



\end{thebibliography}
\end{document}